\documentclass[12pt]{article}
\usepackage{amssymb}
\newcommand{\bee}{\begin{eqnarray}}
\newcommand{\ede}{\end{eqnarray}}
\begin{document}
\title{Study of the Spin-weighted Spheroidal Equation in the
Case of $s=1$ \footnote{ Project supported by the National Natural
Science Foundation of China (Grant Nos. 10875018 and 10773002).} }
\author{Yue Sun \footnote{E-mail : sunyue1101@126.com}  £¬ \ Tian Guihua \footnote{ E-mail: tgh-2000@263.net} and  \ Dong Kun \footnote{ E-mail: woailiuyanbin1@126.com}\\
School of Science, Beijing University of Posts And Telecommunications.
\\ Beijing 100876, China}
\date{(Received 11 November 2010}
\maketitle
\begin{abstract}

We present series study of using the method of super-symmetric quantum mechanics($SUSYQM$) solving the spin-weighted spheroidal wave equation.
In this paper, we obtain the first four terms of super-potential of the spin-weighted spheroidal wave equation in the case of s=1. These results
may help summary the general form for the n-th term of the super-potential£¬which is proved correct by means of induction. We finally compute
the ground eigenvalues and ground
eigenfunction. All the results may be of significative for studies of electromagnetic radiation processes near rotating black holes and compute radiation reaction in curved space-time.\\
\texttt{Keywords}: spheroidal wave equation, supersymmetric quantum mechanics, super-potential, the ground eigenvalue and eigenfunction
\end{abstract}

\section{Introduction }
  The spin-weighted spheroidal functions are first defined by Teukolsky in the study of the perturbations of the Kerr black hole, which are
  indispensable in study of quantum field theory in curved space-time and black hole perturbation theory, etc. Its differential equations
  are$^{[1]-[3]}$
\begin{equation}
 \left[\frac{1}{\sin
\theta}\frac{d}{d\theta}\left(\sin \theta \frac{d}{d
\theta}\right)+s+ a^{2}\omega ^2\cos ^2 \theta -2as\omega \cos
\theta -\frac{\left(m+s\cos \theta\right)^{2}}{\sin ^2
\theta}+E\right]\Theta(\theta)=0,\label{original eq}
\end{equation}
where $E$ is an eigenvalue to be found by imposing appropriate boundary conditions and the parameter $s$, which is the spin-weight of the
perturbation fields could be $s=0, \pm\frac12, \pm1, \pm2$  correspond to the scalar, neutrino, electromagnetic and gravitational perturbations
respectively. And the boundary conditions requires $\Theta$ is finite at $\theta=0,\
 \pi$. In series papers, the perturbation methods in the super-symmetric quantum mechanics($SUSYQM$)
 are used to study the spheroidal equations and many useful results are obtained.$^{[4]-[14]}$
 Here we continue to study the electromagnetic one, which is the spin-weighted equations in the case of
 $s=1$.

 As done before, defining $\Theta(\theta)=\frac{\Psi(\theta)}{\sqrt{sin\theta}}$ and introducing an artificial small parameter$\beta=a\omega$ then Eq.(1)
 leads$^{[4]-[14]}$
\begin{eqnarray}
\frac{d^2\Psi}{d\theta^2}+\left[\frac{1}{4}+s+\beta^{2}\cos ^2
\theta -2s\beta \cos\theta-\frac{(m+s\cos\theta)^2-\frac{1}{4}}{\sin
^2\theta}+E\right]\Psi=0\label{new eq}
\end{eqnarray}
Thus, the corresponding  boundary conditions being $\Psi|_{\theta=0}=\Psi|_{\theta=\pi}=0$. The potential in Eq.(\ref{new eq}) is
\begin{eqnarray}
&&V(\theta,\beta,s)=-\left[\frac{1}{4}+s+\beta^{2}\cos ^2 \theta
-2s\beta \cos\theta-\frac{(m+s\cos\theta)^2-\frac{1}{4}}{\sin
^2\theta}\right]\label{potential},
\end{eqnarray}
Super potential $W$ is an important concept in $SUSYQ$, and it can be determined by potential with the relation as$^{[15]}$
\begin{equation}
W^2-W'=V(\theta,\beta,s)-E_0.\label{potential and super relation}
\end{equation}
And the ground state wave function $\Psi_0$ can be obtained from the super-potential by
\begin{eqnarray}
\Psi_0&=&N\exp\left[-\int Wdz\right]\label{groung function and
super-potential}.
\end{eqnarray}
Both the Eq.(\ref{potential and super relation}) and Eq.(\ref{new eq}) are both difficult to solve so that we look for the approximation
methods. In the paper, we will resolve the spin-weighted spheroidal equations by expanding the super-potential $W$ as a Taylor series in powers
of $\beta$ and the ground eigenvalue $E_0$ can be also expanded by $\beta$$^{[4]-[14]}$
\begin{eqnarray}
W&=&\sum_{n=0}^{\infty}W_{n}\beta^n\label{super-potential expansion}, \\E_0&=&\sum_{n=0}^{\infty}E_{0,n;m}\beta^n\label{energy expansion},
\end{eqnarray}
where $n$ refers to the nth-order item of the series expansion, and $m$ in $E_{0,n;m}$ is referred to the parameter $m$ in Eq.(\ref{new eq}) and
the index $0$ means belonging to the ground state energy. Substituting Eqs.(\ref{super-potential expansion})and (\ref{energy expansion}) into
Eq.(\ref{potential and super relation}), one could  easily obtain
\begin{eqnarray}
W'_0-W_0^2&=&E_{0,0;m}+s+\frac14-\frac{(m+s\cos\theta)^2-\frac{1}{4}}{\sin ^2\theta}\equiv f_0(\theta)\nonumber\\
W'_1-2W_0W_1&=&E_{0,1;m}-2s\cos\theta\equiv f_1(\theta)\nonumber\\
W'_2-2W_0W_2&=&E_{0,2;m}+\cos^2\theta+W_1^2\equiv f_2(\theta)\nonumber\\
W'_{n}-2W_{0}W_{n}&=&E_{0,n;m}+\sum_{k=1}^{n-1}W_{k}W_{n-k}\equiv
f_n(\theta)\ \ \ (n\geq3)\label{f_n define}
\end{eqnarray}
When s =1 , it is straight forward to obtain$^{[12]}$
\begin{eqnarray}
E_{0,0;m}&=&m^2+m-2\nonumber\\
W_0&=&-\frac{1+(m+\frac12)\cos\theta}{sin\theta}
\end{eqnarray}
From Eq.(\ref{f_n define}) it is easy to see that
\begin{eqnarray}
W_{n}(\theta)=(\tan\frac{\theta}{2})^{-2}sin^{-2m-1}\theta A_{n}(\theta)=(1+cos\theta)^2sin^{-2m-3}\theta
A_{n}(\theta)\label{A-n,W-n relation}
\end{eqnarray}
where,
 \begin{eqnarray}
A_{n}(\theta)=\int f_n(\theta)(1-cos\theta)^{2}sin^{2m-1}\theta
d\theta \label{A-n eqn}.
\end{eqnarray}
We introduce the following formulae to simplify calculation in Eqs.(\ref{A-n,W-n relation})-(\ref{A-n eqn})$^{[15]}$
\begin{eqnarray}
P(2m+1,\theta)=\int{sin^{2m+1}\theta}d\theta=-\frac{\cos\theta}{2m+2}\left[\sum_{k=0}^{m}I(2m+1,k)sin^{2m-2k}\theta\right]\label{defination P}
\end{eqnarray}
where,
\begin{equation}
I(2m+1,k)=\left\{\begin{array}{ll}
\frac{(2m+2)(2m)\cdot\cdot\cdot(2m+2-2k)}{(2m+1)(2m-1)\cdot\cdot\cdot(2m-2k+1)} & \textrm{if $k\geq0$}\\
0 &\textrm {if $k<0$}\\
\end{array}
\right.
\end{equation}
Here we give some recurrence relations between $P(2m+2n+1,\theta)$ and $P(2m-1,\theta)$ for later use. Assuming
\begin{eqnarray}
P(2m+2n+1,\theta)=P_1(2m+2n+1,\theta)+P_2(2m+2n+1,\theta),
\end{eqnarray}
where,
\begin{eqnarray}
P_1(2m+2n+1,\theta)&=&\frac{-\cos\theta}{2m+2n+2}\left[\sum_{k=0}^{n}I(2m+2n+1,k)sin^{2m+2n-2k}\theta\right]\\
P_2(2m+2n+1,\theta)&=&\frac{-\cos\theta}{2m+2n+2}\left[\sum_{k=n+1}^{m+n}I(2m+2n+1,k)sin^{2m+2n-2k}\theta\right]\nonumber\\
&=&\frac{-\cos\theta}{2m+2n+2}\left[\sum_{p=0}^{m-1}I(2m+2n+1,p+n+1)sin^{2m-2p-2}\theta\right]\nonumber\\
&=&\frac{(2m-1)I(2m+2n+1,n+1)}{2m+2n+2}P(2m-1,\theta)
\end{eqnarray}
So $P_1(2m+2n+1,\theta)$ contains the item like $\sin^{2m+2j}\theta$ with $j=n-k\geq0$, which is the convergent item at $0,\pi$ and
$P_2(2m+2n+1,\theta)$ can be expressed by $P(2m-1,\theta)$. Hence $P(2m+2n+1,\theta)$ can be rewritten as
\begin{eqnarray}
P(2m+2n+1,\theta)&=&\frac{-\cos\theta}{2m+2n+2}\sum_{k=0}^{n}I(2m+2n+1,k)sin^{2m+2n-2k}\theta\nonumber\\
&&+\frac{(2m-1)I(2m+2n+1,n+1)}{2m+2n+2}P(2m-1,\theta)\label{P(2m+2n+1)P(2m-3)relation}
\end{eqnarray}
By using Eq.(\ref{P(2m+2n+1)P(2m-3)relation}), the calculation of $A_1$ could be further simplified to be
\begin{eqnarray}
A_{1}(\theta)&=&\int (E_{0,1;m}-2cos\theta)(1-cos\theta)^{2}sin^{2m-1}\theta d\theta\nonumber\\
     &=&(2E_{0,1;m}+4)P(2m-1,\theta)-(E_{0,1;m}+4)P(2m+1,\theta)\nonumber\\
&&-\frac{E_{0,1;m}+2}{m}sin^{2m}\theta+\frac{sin^{2m+2}\theta}{m+1}\nonumber\\
&=&\left(2E_{0,1;m}+4-\frac{2m(E_{0,1;m}+4)}{2m+1}\right)*P(2m-1,\theta)\nonumber\\
&&-\frac{E_{0,1;m}+2}{m}sin^{2m}\theta+\frac{sin^{2m+2}\theta}{m+1}+\frac{E_{0,1;m}+4}{2m+1}\cos\theta sin^{2m}\theta\label{A-1 eqn}
\end{eqnarray}
From Eqs.(\ref{A-n,W-n relation}),(\ref{defination P}) the term $P(2m-1,\theta)$ in $A_1(\theta)$ would make $W_1(\theta)$ blow up at the
boundaries $\theta=0,\ \pi$, that is to say unless the coefficient of the divergent term $P(2m-1,\theta)$ in $A_1$ is zero, the eigen-function
$\Psi(\theta)$ could not finite at the boundaries. So $ 2E_{0,1;m}+4-\frac{2m(E_{0,1;m}+4)}{2m+1}=0 $, that is
\begin{equation}
E_{0,1;m}=-\frac{2}{m+1}\label{E-1}
\end{equation}
Equ.(19) can simplify $A_1$ in the concise form
\begin{eqnarray}
A_{1}(\theta)&=&\frac{sin^{2m}\theta}{m+1}[\sin^2\theta-2+2\cos\theta]\label{A-1}
\end{eqnarray}
Owing to Eq.(\ref{A-n,W-n relation}), it is easy to obtain the $W_1(\theta)$
\begin{eqnarray}
W_{1}(\theta)&=&A_{1}(\theta)(1+cos\theta)^2(sin\theta)^{-2m-3}=-\frac{1}{m+1}\sin\theta
\end{eqnarray}
By the same calculation  process,  one could obtain $W_{2}(\theta)$ and $W_{3}(\theta)$ as
\begin{eqnarray}
W_{2}(\theta)&=&\frac{m(m+2)\cos\theta\sin\theta}{(2m+3)(m+1)^2}-\frac{m(m+2)\sin\theta}{(2m+3)(m+1)^3}\\
W_{3}(\theta)&=&-\frac{2m\cos\theta \sin\theta}{(m+1)^4(2m+3)}+\frac{2m\sin\theta}{(m+1)^{5}(2m+3)}-\frac{m\sin^3\theta}{(m+1)^3(2m+3)}
\end{eqnarray}
and the corresponding $E_{0,2;m}$ and $E_{0,3;m}$ are
\begin{eqnarray}
E_{0,2;m}&=&-\frac{m^3+7m^2+11m+3}{(m+1)^3(2m+3)}\nonumber\\
E_{0,3;m}&=&-\frac{4m^2(m+2)}{(m+1)^5(2m+3)}
\end{eqnarray}
\section{The general formula of n-th $W_n$ for the super-potential}
In this section, we summarize the general formula for $W_n$ from the results of $W_1$ to $W_3$ as following
\begin{equation}
W_{n}(\theta)=\cos\theta\sum_{k=1}^{[\frac{n}{2}]}a_{n,k}\sin^{2k-1}\theta+\sum_{k=1}^{[\frac{n+1}{2}]}b_{n,k}\sin^{2k-1}\theta\label{Wn-general}
\end{equation}
where the coefficient $a_{n,k}$ and $b_{n,k}$ need to satisfy
\begin{equation}
\begin{array}{ll}
a_{n,k}=0 & \textrm{if $k<1$ or $k>[\frac{n}{2}]$}\\
b_{n,k}=0 & \textrm{if $k<1$ or $k>[\frac{n+1}{2}]$}\label{ab}
\end{array}
\end{equation}
Eq.(\ref{Wn-general}) is a form different from the super-potential $W_n$ in Ref.(9), that is
\begin{equation}
W_{n}(\theta)=\sin\theta\sum_{k=0}^{n-1}\hat{a}_{n,k}\cos^{2k+1}\theta, n=1,2,\cdots
\end{equation}

The next work is to prove its correctness by mathematical induction. Obviously $W_N$ for $N=1,2,3$ satisfies Eq.(\ref{Wn-general}). Assuming
that when $N<n$, $W_N$ meet the general form, we need to verify $W_n(\theta)$ also satisfy the general formula. Back to Eq.(\ref{f_n define}),
taking $W_{k}(\theta)$ and $W_{n-k}(\theta)$ into $\sum_{k=1}^{n-1}W_{k}W_{n-k}$, that is
\begin{eqnarray}
\sum_{k=1}^{n-1}W_{k}W_{n-k}=\sum_{p=2}^{[\frac{n}{2}]+1}h_{n,p}\sin^{2p-2}\theta+\sum_{p=2}^{[\frac{n}{2}]+1}g_{n,p}\sin^{2p-2}\theta\cos\theta
\end{eqnarray}
where $h_{n,p}$ and $g_{n,p}$ are constant coefficients:$^{[12]}$
\begin{eqnarray}
h_{n,p}&=&\sum_{k=1}^{n-1}\sum_{j=1}^{p-1}(a_{k,p-j}a_{n-k,j}
-a_{k,p-1-j}a_{n-k,j}+b_{k,p-j}b_{n-k,j})\nonumber\\
g_{n,p}&=&
\sum_{k=1}^{n-1}\sum_{j=1}^{p-1}(a_{k,p-j}b_{n-k,j}+b_{k,p-j}a_{n-k,j})
\end{eqnarray}
In order to simplify the calculation process,  the maximum limit of $p$ can be written uniformly as $[\frac n2]+1$, because of Eq.(\ref{ab}), we
can get
\begin{equation}
\begin{array}{ll}
h_{n,p}=0, & \textrm{if $p<2$ or $p>[\frac{n}{2}]+1$}\\
g_{n,p}=0, & \textrm{if $p<2$ or
$p>[\frac{n+1}{2}]$}\label{C_{n,p,g_{n,p}}}
\end{array}
\end{equation}
 Taking
$f_n(\theta)=E_{0,n;m}+\sum_{k=1}^{n-1}W_{k}W_{n-k}$ into Eq.(\ref{A-n eqn})£¬ $A_n(\theta)$ can be written as
\begin{eqnarray}
&&A_{n}(\theta)\nonumber\\
&=&\int\left[E_{0,n;m}+\sum_{p=2}^{[\frac{n}{2}]+1}h_{n,p}\sin^{2p-2}\theta+\cos\theta
\sum_{p=2}^{[\frac{n}{2}]+1}g_{n,p}\sin^{2p-2}\theta\right]
(1-cos\theta)^{2}sin^{2m-1}\theta d\theta\nonumber\\
&=& E_{0,n;m}\left[2P(2m-1,\theta)-P(2m+1,\theta)-\frac{\sin^{2m}\theta}{m}\right]\nonumber\\
&&+\sum_{p=2}^{[\frac{n}{2}]+1}\left[-\frac{g_{n,p}\sin^{2m+2p}\theta}{2m+2p}+\frac{(g_{n,p}-h_{n,p})\sin^{2m+2p-2}\theta}{m+p-1}\right]\nonumber\\
&&+\sum_{p=2}^{[\frac{n}{2}]+1}\left[2(h_{n,p}-g_{n,p})P(2m+2p-3,\theta)+(2g_{n,p}-h_{n,p})P(2m+2p-1,\theta)\right]
\end{eqnarray}
We simplify the above result according to Eq.(\ref{P(2m+2n+1)P(2m-3)relation}), that is
\begin{eqnarray}
&&A_{n}(\theta)\nonumber\\
&=&P(2m-1,\theta)\times b_1+E_{0,n;m}\left[-\frac{\sin^{2m}\theta}{m}+\cos\theta\frac{\sin^{2m}\theta}{2m+1}\right]\nonumber\\
&&+\sum_{p=2}^{[\frac{n}{2}]+1}\left[-\frac{g_{n,p}\sin^{2m+2p}\theta}{2m+2p}+\frac{(g_{n,p}-h_{n,p})\sin^{2m+2p-2}\theta}{m+p-1}\right]\nonumber\\
&&+\sum_{p=2}^{\frac{[n]}{2}+1}(2h_{n,p}-2g_{n,p})\left[\frac{-\cos\theta}{2m+2p-2}\sum_{j=0}^{p-2}I(2m+2p-3,p-2-j)sin^{2m+2j}\theta\right]\nonumber\\
&&+\sum_{p=2}^{\frac{[n]}{2}+1}(2g_{n,p}-h_{n,p})\left[\frac{-\cos\theta}{2m+2p}\sum_{j=0}^{p-1}I(2m+2p-1,p-1-j)sin^{2m+2j}\theta\right]\label{orgAn}
\end{eqnarray}
where $b_1$ is the coefficient of the term $P(2m-1,\theta)$
\begin{eqnarray}
b_1&=&\left[\frac{2(m+1)E_{0,n;m}}{2m+1}+(2m-1)\sum_{p=2}^{[\frac{n}{2}]+1}\frac{h_{n,p}-g_{n,p}}{m+p-1}I(2m+2p-3,p-1)\right]\nonumber\\
&&+(2m-1)\sum_{p=2}^{[\frac{n}{2}]+1}\left[\frac{2g_{n,p}-h_{n,p}}{2m+2p}I(2m+2p-1,p)\right]\label{E0,n;m=0}
\end{eqnarray}
As similar analysis of $A_1(\theta)$, $W_n(\theta)$ blows up at the boundaries $\theta=0,\ \pi$ when $b_1\neq0$ we must set the coefficient
$b_1$ of the term $P(2m-1,\theta)$ in $A_n$ zero in order that the eigen-function $\Psi_0(\theta)$ could finite at the boundaries. Thus $b_1=0$
gives
\begin{eqnarray}
&&E_{0,n;m}= -\frac{(2m+1)(2m-1)}{2(m+1)}
\bigg[\sum_{p=2}^{[\frac{n}{2}]+1}\frac{h_{n,p}-g_{n,p}}{m+p-1}I(2m+2p-3,p-1)\bigg]\nonumber\\
&& \ \ \ \ \ \ \ \ \ \ \ \ \ -\frac{(2m+1)(2m-1)}{2(m+1)}
\bigg[\sum_{p=2}^{[\frac{n}{2}]+1}\frac{2g_{n,p}-h_{n,p}}{2m+2p}I(2m+2p-1,p)\bigg]\label{E_0,n;m}
\end{eqnarray}
By the result of the Eq.(\ref{E_0,n;m}), $A_n(\theta)$ can be simplified as $A_{n}(\theta)=R_n(\theta)+\cos T_n(\theta)$, where
\begin{eqnarray}
R_n(\theta)&=&-\frac{E_{0,n;m}}{m}\sin^{2m}\theta+\sum_{p=2}^{[\frac{n}{2}]+1}\left[\frac{(g_{n,p}-h_{n,p})\sin^{2m+2p-2}\theta}{m+p-1}-\frac{g_{n,p}\sin^{2m+2p}\theta}{2m+2p}\right]\nonumber\\
&=&\sum_{p=0}^{[\frac{n}{2}]+1}R_{n,p}\sin^{2m+2p}\theta\label{Rn}
\end{eqnarray}
with $R_{n,0}=-\frac{E_{0,n;m}}{m}$ and $R_{n,1}=\frac{g_{n,2}-h_{n,2}}{m+1}$, and when $p\geq2$, $R_{n,p}$ is
\begin{eqnarray}
R_{n,p}&=&\frac{2g_{n,p+1}-2h_{n,p+1}-g_{n,p}}{2m+2p}
\end{eqnarray}
and
\begin{eqnarray}
T_n(\theta)&=&\frac{E_{0,n;m}}{2m+1}\sin^{2m}\theta+\sum_{p=2}^{\frac{[n]}{2}+1}\left[\frac{g_{n,p}-h_{n,p}}{m+p-1}\sum_{j=0}^{p-2}I(2m+2p-3,p-2-j)sin^{2m+2j}\theta\right]\nonumber\\
&&+\sum_{p=2}^{\frac{[n]}{2}+1}\left[\frac{h_{n,p}-2g_{n,p}}{2m+2p}\sum_{j=0}^{p-1}I(2m+2p-1,p-1-j)sin^{2m+2j}\theta\right]\nonumber\\
&=&\frac{E_{0,n;m}}{2m+1}\sin^{2m}\theta\nonumber\\
&&+\sum_{j=0}^{\frac{[n]}{2}-1}\sum_{p=j+2}^{\frac{[n]}{2}+1}\left[\frac{g_{n,p}-h_{n,p}}{m+p-1}I(2m+2p-3,p-2-j)sin^{2m+2j}\theta\right]\nonumber\\
&&+\sum_{j=0}^{\frac{[n]}{2}}\sum_{p=j+1}^{\frac{[n]}{2}+1}\left[\frac{h_{n,p}-2g_{n,p}}{2m+2p}I(2m+2p-1,p-1-j)sin^{2m+2j}\theta\right]\nonumber\\
&=&\sum_{j=0}^{[\frac{n}{2}]}T_{n,j}\sin^{2m+2j}\theta\label{Tn}
\end{eqnarray}
with
\begin{eqnarray}
T_{n,0}&=&\frac{E_{0,n;m}}{2m+1}+\sum_{p=2}^{\frac{[n]}{2}+1}\left[\frac{g_{n,p}-h_{n,p}}{m+p-1}I(2m+2p-3,p-2)\right]\nonumber\\
&&+\sum_{p=1}^{\frac{[n]}{2}+1}\left[\frac{h_{n,p}-2g_{n,p}}{2m+2p}I(2m+2p-1,p-1)\right]
\end{eqnarray}
and when $j\geq1$, $T_{n,j}$ being abbreviated as
\begin{eqnarray}
T_{n,j}&=&\sum_{p=j+2}^{\frac{[n]}{2}+1}\left[\frac{g_{n,p}-h_{n,p}}{m+p-1}I(2m+2p-3,p-2-j)\right]\nonumber\\
&&+\sum_{p=j+1}^{\frac{[n]}{2}+1}\left[\frac{h_{n,p}-2g_{n,p}}{2m+2p}I(2m+2p-1,p-1-j)\right]
\end{eqnarray}
Due to Eq.(\ref{ab}), one can get
\begin{equation}
\begin{array}{ll}
R_{n,j}=0, & \textrm{if $j<-1$ or $j>[\frac{n+1}{2}]$}\\
T_{n,j}=0, & \textrm{if $j<-1$ or $j>[\frac{n}{2}]$}\label{RnTn}
\end{array}
\end{equation}
According to Eq.(\ref{A-n,W-n relation}),
\begin{eqnarray}
W_n(\theta)=\left[R_n(\theta)+\cos\theta T_n(\theta)\right](1+\cos\theta)^2\sin^{-2m-3}\theta=X_n+\cos\theta Y_n\label{wn-xn-yn}
\end{eqnarray}
with
\begin{eqnarray}
X_n&=&\left[(2R_n+2T_n)-(R_n+2T_n)sin^{2}\theta\right]\sin^{-2m-3}\theta\nonumber\\
&=&\sum_{j=0}^{[\frac{n}{2}]+1}\bigg[(2R_{n,j}+2T_{n,j})sin^{2j-3}\theta-(R_{n,j}+2T_{n,j})sin^{2j-1}\theta\bigg]\nonumber\\
&=&\sum_{j=-1}^{[\frac{n}{2}]+1}X_{n,j}sin^{2j-1}\theta,
\end{eqnarray}
where
\begin{eqnarray}
X_{n,j}&=&2R_{n,j+1}+2T_{n,j+1}-R_{n,j}-2T_{n,j}\nonumber\\
&=&\sum_{p=j+1}^{\frac{[n]}{2}+1}\frac{[(2m+2p)h_{n,p}-2g_{n,p}]I(2m+2p-1,p-1-j)}{(m+j+1)(2m+2p)(2m+2p-2)}+\frac{g_{n,j}}{2m+2j}
\end{eqnarray}
 And $Y_n$ can be expressed as
\begin{eqnarray}
Y_n&=&\left[(2R_n+2T_n)-T_nsin^{2}\theta\right]\sin^{-2m-3}\theta\nonumber\\
&=&\sum_{j=-1}^{[\frac{n}{2}]+1}\left[2R_{n,j+1}+2T_{n,j+1}-T_{n,j}\right]sin^{2j-1}\theta\nonumber\\
&=&\sum_{j=-1}^{[\frac{n}{2}]+1}Y_{n,j}sin^{2j-1}\theta,
\end{eqnarray}
from Eq.(35) and Eq.(38), we can get
\begin{eqnarray}
Y_{n,j}&=&2R_{n,j+1}+2T_{n,j+1}-T_{n,j}\nonumber\\
&=&-\sum_{p=j+1}^{\frac{[n]}{2}+1}\frac{(m+j)[(2m+2p)h_{n,p}-2g_{n,p}]I(2m+2p-1,p-1-j)}{(m+j+1)(2m+2p)(2m+2p-2)}
\end{eqnarray}
Obviously, in the case of $j=-1,0,$ the items in $X_{n,j}$ and $Y_{n,j}$ of $A_n(\theta)$ may cause $W_n(\theta)$ divergent, then we need to
checking whether these items are exist
\begin{eqnarray}
X_{n,-1}&=&2R_{n,0}+2T_{n,0}=0\nonumber\\
X_{n,0}&=&2R_{n,1}+2T_{n,1}-R_{n,0}-2T_{n,0}=0\nonumber\\
Y_{n,-1}&=&2R_{n,0}+2T_{n,0}=0\nonumber\\
Y_{n,0}&=&2R_{n,1}+2T_{n,1}-T_{n,0}=0
\end{eqnarray}
Hence, all the items which can make $W_n(\theta)$ divergent are not exist, so that $W_n$ could be convergent in the boundary region.
$W_n(\theta)$ now becomes
\begin{eqnarray}
W_n(\theta)&=&\sum_{j=1}^{[\frac{n}{2}]+1}X_{n,j}sin^{2j-1}\theta+\cos\theta
\sum_{j=1}^{[\frac{n}{2}]+1}Y_{n,j}sin^{2j-1}\theta\nonumber\\
&=&\sum_{l=1}^{[\frac{n}{2}]+1}b_{n,l}sin^{2l-1}\theta+\cos\theta
\sum_{l=1}^{[\frac{n}{2}]+1}a_{n,l}sin^{2l-1}\theta\label{W*}
\end{eqnarray}
where
\begin{eqnarray}
a_{n,l}&=&Y_{n,j}=-\sum_{p=j+1}^{\frac{[n]}{2}+1}\frac{(m+j)[(2m+2p)h_{n,p}-2g_{n,p}]I(2m+2p-1,p-1-j)}{(m+j+1)(2m+2p)(2m+2p-2)}\nonumber\\
b_{n,l}&=&X_{n,j}=\sum_{p=j+1}^{\frac{[n]}{2}+1}\frac{[(2m+2p)h_{n,p}-2g_{n,p}]I(2m+2p-1,p-1-j)}{(m+j+1)(2m+2p)(2m+2p-2)}+\frac{g_{n,j}}{2m+2j}
\end{eqnarray}
Note that the maximum limit of $l$ in Eq.(\ref{W*}) is differ with the general formula Eq.(\ref{Wn-general}). By the relation Eq.(\ref{RnTn}) we
could analysis the maximum limit of $l$ with the parity of $n$. When $n$ is even, $l$ gets the maximum $l=[\frac{n}{2}]+1=\frac{n}{2}+1$, then
\begin{eqnarray}
a_{n,\frac{n}{2}+1}&=&2R_{n,\frac{n}{2}+2}+2T_{n,\frac{n}{2}+2}-T_{n,\frac{n}{2}+1}=0\nonumber\\
b_{n,\frac{n}{2}+1}&=&2R_{n,\frac{n}{2}+2}+2T_{n,\frac{n}{2}+2}-R_{n,\frac{n}{2}+1}-2T_{n,\frac{n}{2}+1}=0
\end{eqnarray}
but when $l=[\frac{n}{2}]=\frac{n}{2}$, $a_{n,\frac{n}{2}}$ and $b_{n,\frac{n}{2}}$ are both nonzero. Hence under the condition $n$ is even
$W_n(\theta)$ becomes
\begin{eqnarray}
W_n(\theta)&=&\cos\theta
\sum_{l=1}^{[\frac{n}{2}]}a_{n,l}sin^{2l-1}\theta
+\sum_{l=1}^{[\frac{n}{2}]}b_{n,l}sin^{2l-1}\theta,\label{Wn-even}
\end{eqnarray}
When $n$ is odd, $l$ gets the maximum
   $l=[\frac{n}{2}]+1=\frac{n+1}{2}$, then
\begin{eqnarray}
a_{n,\frac{n+1}{2}}&=&2R_{n,\frac{n+3}{2}}+2T_{n,\frac{n+3}{2}}-T_{n,\frac{n+1}{2}}=0\nonumber\\
b_{n,\frac{n+1}{2}}&=&2R_{n,\frac{n+3}{2}}+2T_{n,\frac{n+3}{2}}-R_{n,\frac{n+1}{2}}-2T_{n,\frac{n+1}{2}}\neq0
\end{eqnarray}
and when $k=[\frac{n}{2}]=\frac{n-1}{2}$, $a_{n,\frac{n-1}{2}}$ and $b_{n,\frac{n-1}{2}}$ are nonzero. While when n is odd $W_n(\theta)$ has the
new form
\begin{eqnarray}
W_n(\theta)&=&\cos\theta
\sum_{l=1}^{[\frac{n}{2}]}a_{n,l}sin^{2l-1}\theta
+\sum_{l=1}^{[\frac{n+1}{2}]}b_{n,l}sin^{2l-1}\theta,\label{Wn-odd}
\end{eqnarray}
with the help $[\frac n2]+1=[\frac{n+1}2]$. Through comparing the Eq.(\ref{Wn-even}) and Eq.(\ref{Wn-odd}) the super-potential $W_n(\theta)$ can
be rewritten as
\begin{eqnarray}
W_n(\theta)&=&\cos\theta \sum_{l=1}^{[\frac{n}{2}]}a_{n,l}sin^{2l-1}\theta+\sum_{l=1}^{[\frac{n+1}{2}]}b_{n,l}sin^{2l-1}\theta
\end{eqnarray}
This completes the suppose of the general formula $W_n$. That is to say, for any $n\geq1$, $W_n$ satisfy the form of Eq.(\ref{Wn-general}).
\section{The ground eigenfunction }
From the super-potential
\begin{eqnarray}
 W&=&W_0+\sum_{n=1}^{\infty}\left[\cos\theta\sum_{k=1}^{[\frac{n}{2}]}a_{n,k}\sin^{2k-1}\theta+\sum_{k=1}^{[\frac{n+1}{2}]}b_{n,k}\sin^{2k-1}\theta\right]\beta^n
\end{eqnarray}
the ground eigenfunction becomes
\begin{eqnarray}
\Psi_0&=&N\exp\left[-\int Wdz\right]\nonumber\\
        &=&N\exp\left[\int \frac{1+(m+\frac12)\cos\theta}{sin\theta} d\theta-\sum_{n=1}^{\infty}\beta^n
        \int (\cos\theta\sum_{k=1}^{[\frac{n}{2}]}a_{n,k}+\sum_{k=1}^{[\frac{n+1}{2}]}b_{n,k})\sin^{2k-1}\theta d\theta\right] \nonumber\\
        &=&N(1-\cos\theta)\sin^{m+\frac{1}{2}}\theta\exp\left[-\sum_{n=1}^{\infty}\beta^n
        \left(\sum_{k=1}^{[\frac{n}{2}]}a_{n,k}\frac{\sin^{2k}\theta}{2k}+\sum_{k=1}^{[\frac{n+1}{2}]}b_{n,k}P(2k-1,\theta)\right)\right]\nonumber\\
\end{eqnarray}
and the ground eigenvalue is
\begin{eqnarray}
E_{0,n;m}&=&E_{0,0;m}+\sum_{n=1}^{\infty}\beta^nE_{0,n;m} =m^2+m-2+\sum_{n=1}^{\infty}E_{0,n;m}\beta^n
\end{eqnarray}
$E_{0,n;m}$ has been obtained by Eq.(\ref{E_0,n;m}).

\end{document}